\documentclass[12pt]{JHEP} % 10pt is ignored!

\title{Global Analysis of Solar Neutrino Oscillations Including SNO CC
Measurement}

\author{John N. Bahcall\\
  School of Natural Sciences, Institute for Advanced Study, Princeton,
  NJ 08540\\
	E-mail: \email{jnb@ias.edu}}
\author{M. C. Gonzalez-Garcia \\
  Theory Division, CERN, CH-1211, Geneva 23, Switzerland,\\
  C.N. Yang Institute for Theoretical Physics\\
State University of New York at Stony Brook\\
Stony Brook,NY 11794-3840, USA,\\
and Instituto de F\'\i sica Corpuscular, Universitat de Val\`encia
  -- C.S.I.C. \\ Edificio Institutos de Paterna, Apt 22085, 46071
  Val\`encia, Spain\\
	E-mail: \email{concha@thwgs.cern.ch}}
\author{Carlos Pe\~na-Garay\\
        Instituto de F\'\i sica Corpuscular, Universitat de Val\`encia
  -- C.S.I.C. \\ Edificio Institutos de Paterna, Apt 22085, 46071
  Val\`encia, Spain\\
       E-mail: \email{penya@ific.uv.es}}
\preprint{CERN-TH/2001-165}	% OR: \preprint{Aaaa/Mm/Yy\\Aaa-aa/Nnnnnn}
\abstract{For active and sterile neutrinos, we present the globally
allowed solutions for two neutrino oscillations. We include the SNO CC
measurement and all other relevant solar neutrino and reactor
data. Five active neutrino oscillation solutions (LMA, LOW, SMA, VAC,
and Just So$^2$) are currently allowed at $3\sigma$; three sterile
neutrino solutions (Just So$^2$, SMA, and VAC) are allowed at
$3\sigma$. The goodness of fit is satisfactory for all eight
solutions. We also investigate the robustness of the allowed solutions
by carrying out global analyses with and without: 1) imposing solar
model constraints on the $^8$B neutrino flux, 2) including the
Super-Kamiokande spectral energy distribution and day-night data, 3)
including a continuous mixture of active and sterile neutrinos, 4)
using an enhanced CC cross section for deuterium (due to radiative
corrections), and 5) an optimistic, hypothetical reduction by a factor
of three of the error of the SNO CC rate. For every analysis strategy
used in this paper, the most favored solutions all involve large
mixing angles: LMA, LOW, or VAC.  The favored solutions are robust,
but the existence at $3\sigma$ of individual sterile solutions and the
active Just So$^2$ solution is sensitive to the analysis assumptions.}


\begin{document} 
\input psfig

\section{Introduction}
The epochal Sudbury Neutrino Observatory (SNO)
measurement~\cite{sno2001} of the CC rate for solar neutrino
absorption by deuterium provides an important new constraint on the
allowed neutrino oscillation solutions.  We present in this paper
solutions for the globally allowed regions that include, in addition
to the SNO CC measurement, the available data from the
Chlorine~\cite{chlorine}, Gallium~\cite{gallex,sage,gno}, and
Super-Kamiokande~\cite{superk} experiments.  We emphasize tests of the
robustness of the global solutions to different analysis strategies.

The most radical test we make is to determine the allowed solutions if
only the total rates of the solar neutrino experiments are used,
ignoring the beautiful data from the Super-Kamiokande measurements of
the spectral energy distribution and the day-night variations.  This
test is motivated by the fact that the Super-Kamiokande day and night
recoil energy spectra provide $38$ different data points, while the
rates of the chlorine experiment, the two gallium experiments, and SNO
provide together only $4$ data points. Some authors have questioned
whether $\chi^2$ fits to the complete data set overweight the
Super-Kamiokande data because of the large number of spectral energy
bins~\cite{Creminelli:2001ij}.

We take both sides of the somewhat philosophical question of whether
the $^8$B neutrino flux should be allowed to vary without considering
constraints from the solar model.  We find self-consistent global
solutions both with and without imposing the solar model
constraints. 

We require (cf. \cite{bks2001}) that the $^8$B neutrino flux be
treated the same way in all aspects of the calculation, either
completely free when evaluating both the rates and the energy spectrum
or constrained everywhere by the solar model predictions.  We also
require that the $^8$B neutrino flux be identical in all parts of the
calculation. Finally, we avoid double counting the Super-Kamiokande
rate measurement; we do not include the measured flux normalization
both in the rate measurement and in the spectral energy
distribution. 

We treat sterile neutrinos on the same basis as active neutrinos,
which implies that our $\chi^2$ plots are made in terms of three free
parameters : $\Delta m^2$, $\tan^2\theta$, and a third parameter,
$\cos^2\eta$, which defines the active-sterile admixture.  We carry
out calculations for a variety of values of $\cos^2\eta$, but, with
one exception that is discussed in Section~\ref{sec:ratesonly} and
Section~\ref{subsec:mixtures}, the absolute minimum in $\chi^2$ always lies in
the active neutrino plane. We also describe how the allowed regions
are affected if one rules out {\it a priori} the possibility that
sterile neutrinos exist and therefore constructs the allowed regions
in $\chi^2$ using, as has usually been done in the past, only two free
parameters: $\Delta m^2$ and $\tan^2\theta$.

Why do all these tests?  There are not yet enough solar neutrino
 experiments to ensure cross checks and redundancy in the
 data. Therefore, different plausible analysis schemes can lead to
 different conclusions.  The existence of a particular allowed region,
 e. g., the SMA solution or the Just So$^2$ solution, may depend upon
 which of several possible plausible analysis schemes are used. Our
 motto is: ``If its not robust, its not believable.''

In section~\ref{sec:calculational}, we summarize the calculational
procedures. Because of the excellent agreement between the predicted
standard solar model flux of $^8$B neutrinos (see ref.~\cite{bp2000},
hereafter BP00) and the combined SNO~\cite{sno2001} and
Super-Kamiokande~\cite{superk} measurement, we include the theoretical
fluxes and their BP00 uncertainties in our standard analysis. We also
use a `level playing field' prescription for evaluating the allowed
active and sterile neutrino solutions. We present in
section~\ref{sec:globalplus} the allowed solutions that exist if all
the solar neutrino data, including the spectral energy distribution
and day night data of Super-Kamiokande, are included in the
analysis. Considering only the total rates in the chlorine, gallium,
Super-Kamiokande, and SNO experiments, we present in
section~\ref{sec:ratesonly} the allowed solutions for this extreme
case. We also investigate the effects of allowing the $^8$B flux to
vary unconstrained by solar model predictions, of using a larger CC
cross section for deuterium (motivated by possible effects of
radiative corrections~\cite{beacom}), and of a hypothetical reduction
by a factor of three in the total quoted error for the SNO CC rate. In
section~\ref{sec:comparisons}, we discuss how some different
approaches to the analysis of solar neutrino data affect the allowed
regions. We pay particular attention in this section to the {\it a
priori} rejection of sterile neutrinos, the role of the measured $^8$B
neutrino flux normalization and the influence of the standard solar
model neutrino flux constraint, and the influence of the day-night
spectral energy data on the allowed sterile neutrino component. We
summarize our results in section~\ref{sec:discussion}.

\section{Calculational Procedures}
\label{sec:calculational}

This section is intended primarily for aficionados
of neutrino oscillation analyses.

We use, unless stated otherwise, the techniques and parameters for the
analysis that we have described elsewhere~\cite{bks2001}
and~\cite{cc2001}. In ref.~\cite{bks2001}, from which we derive our
primary analysis strategy, the focus was on allowed solutions in which
the $^8$B flux was unconstrained by any solar model considerations. In
this paper, we derive and contrast global solutions in which the $^8$B
flux is constrained, or unconstrained, by the standard solar model
uncertainty.

 We first describe in
section~\ref{subsec:defnchi2} the procedure we use for calculating 
 $\chi^2$ for the global analysis and then discuss in
section~\ref{subsec:level} the impartial manner in which we treat
active and sterile neutrinos.

\subsection{Definition of $\chi^2$ for the global analysis}
\label{subsec:defnchi2}

In section~\ref{sec:globalplus}, we determine the allowed range of the
oscillation parameters using the CC event rate measured at SNO, the
Chlorine and Gallium event rates (we use here the weighted averaged
GALLEX/GNO and SAGE rates), and the $2\times 19$ bins of the (1258
day) Super-Kamiokande electron recoil energy spectrum measured
separately during the day and night periods.  In this global analysis
we adopt the prescription described in ref.~\cite{bks2001}. We do not
include here the Super-Kamiokande total rate, since to a large extent
the total rate is represented by the flux in each of the spectral
energy bins.  We define the $\chi^2$ function for the global analysis
as:
\begin{equation}
\chi^2_{\rm global}=\sum_{i,j=1,41} (R^{th}_i- R^{exp}_i)
\sigma_{G,ij}^{-2} (R^{th}_j- R^{exp}_j) ,
\end{equation}
where $\sigma_{G,ij}^2 = \sigma_{R,ij}^2 + \sigma_{Sp,ij}^2$.  Here
$\sigma_{R,ij}$ is the corresponding $41 \times 41$ error matrix
containing the theoretical as well as the experimental statistical and
systematic uncorrelated errors for the 41 rates while $\sigma_{Sp,ij}$
contains the assumed fully--correlated systematic errors for the
$38\times 38$ submatrix corresponding to the Super-Kamiokande
day--night spectrum data.  We include here the energy independent
systematic error which is usually quoted as part of the systematic
error of the total rate. The error matrix $\sigma_{R,ij}$ includes
important correlations arising from the theoretical errors of the
solar neutrino fluxes, or equivalently of the solar model parameters.

When considering just the total rates (see
section~\ref{sec:ratesonly}), we adapt the $\chi^2$ definition of
ref.~\cite{fogli,sno4} to include the two different gallium rates and
the new SNO CC event rate. The $\chi^2$ for this case is
\begin{equation}
\chi^2_{\rm Rates}=\sum_{i,j=1,5} (R^{th}_i- R^{exp}_i)
\sigma_{R,ij}^{-2} (R^{th}_j- R^{exp}_j),
\end{equation}
where $R^{th}_i$ is the theoretical prediction of the event rate in
detector $i$ and $R^{exp}_i$ is the measured rate. The error matrix
$\sigma_{ij}$ contains the experimental errors, both systematic and
statistical, as well as the theoretical uncertainties on the solar
neutrino fluxes and the interaction cross sections.  The theoretical
error matrix includes important correlations arising from the
theoretical errors of the solar neutrino fluxes, or equivalently of
the solar model parameters.  Furthermore for the GALLEX/GNO and SAGE
rates, the corresponding theoretical errors for the interaction cross
section are assumed to be fully correlated.  In our statistical
treatment of the rate data, we adapt the analysis of
ref.~\cite{fogli}, with the updated uncertainties and distributions
for neutrino production fractions and the solar matter density given
in ref.~\cite{bp2000} and tabulated in http:www.sns.ias.edu/$\sim$jnb
.

We also present the results of an analysis performed with an
unconstrained $^8$B neutrino flux. In this analysis, the SSM $^8$B
neutrino flux is multiplied by a factor of $f_{\rm B}$, which is fit
to the data.  We minimize $\chi^2$ with respect to $f_{\rm B}$ for
each set of neutrino oscillation parameters; no theoretical error is
included for the $^8$B neutrino flux in the error matrix. We restrict
ourselves to the range $f_{\rm B}<2$. We have checked that there are
no allowed solutions for values of $f_{\rm B} >2$ if the
Super-Kamiokande day-night recoil energy spectra are included.  In the
minimization procedure, the factor $f_{\rm B}$ is required to be the
same for the normalization of each Super-Kamiokande spectral energy
bin, the SNO CC rate, and  the $^8$ B contributions to the
gallium and chlorine experiments.

One of the largest systematic uncertainties in interpreting the SNO CC
measurement results from the uncertainty in the absolute value of the
neutrino absorption cross section on
deuterium~\cite{sno2001,snoten}. We have used in our calculations the
cross sections calculated by Nakamura et al.~\cite{nakamura}, which
are in good agreement with the results of Butler et
al.~\cite{butler}. Most recently, Beacomand Parke~\cite{beacom} have
argued that there may be a $6$\% increase in the cross sections
evaluated by Nakamura et al and Butler et al. due to a combination of
factors represented by a $2$\% correction to $g_A^2$, and two other
factors related to radiative corrections~\cite{towner}. We follow the
SNO collaboration~\cite{sno2001,private} in including the $2$\%
correction from $g_A^2$ in our standard calculations.

In order to test the sensitivity of the global fits to the effects of
radiative corrections, we have repeated all of our calculations with a
CC cross section for SNO that is increased by an additional $4$\%
relative to the standard cross sections. We shall describe these
calculations in the text as having been done with the 'enhanced CC
cross sections.''

For the SNO CC calculation, we have used the resolution function given
in ref.~\cite{sno2001}, which is slightly broader than we have assumed
in our previous analyses.

Following Fogli, Lisi, and Montanino~\cite{lisitan} and de Gouvea,
Friedland, and Murayama~\cite{GFM}, we present our results in terms of
$\tan^2 \theta$ rather than $\sin^2 2\theta$ in order to include
solutions with mixing angles greater than $\pi/4$ (the so-called `dark
side').

\subsection{A level playing field for sterile neutrinos}
\label{subsec:level}

Once the SNO results are included in the analysis, all of the
solutions with sterile neutrinos are relatively poor fits to the totality
of solar neutrino data. Many different regions of the sterile neutrino
parameter space provide comparable fits to the available data.
We have therefore made one significant departure from previous
analysis techniques for two neutrino oscillations: we treat the active
and sterile neutrinos as different aspects of the same two-neutrino
oscillation scheme.  This procedure 'levels the playing field' for
active and sterile neutrinos. The assumption of oscillation into
either all active or all sterile neutrinos describes the limiting
extremes of a continuum in which the oscillation occurs into a linear
combination of active and sterile neutrinos, see discussion in
ref.~\cite{four} of four neutrino oscillations. 

Our theoretical framework contains three free parameters: $\Delta
m^2$, $\tan^2\theta$, and the third parameter, $\cos^2\eta$, which
defines the active-sterile admixture. We focus here only on the two
limiting cases,$\cos^2\eta = 1,0$\,.  Since the parameter
space is three-dimensional, the allowed regions for a given C.L. are
defined as the set of points satisfying the condition
\begin{equation}
    \chi^2_{\rm sol}(\Delta m^2,\theta,\eta)
    -\chi^2_{\rm sol,min}\leq \Delta\chi^2 \mbox{(C.L., 3~d.o.f.)} , 
\label{eqn:chidiff}
\end{equation}
where $\Delta\chi^2(\mbox{C.L., 3~d.o.f.}) = 6.25$, $7.81$, $11.34$, and
$14.16$ for C.L.~= 90\%, 95\%, 99\% and 99.73\% ($3\sigma$)
respectively, and $\chi^2_{\rm sol,min}$ is the global minimum in the
three-dimensional space. A similar procedure is used in searching for
allowed solutions for three and four neutrino
oscillations~\cite{four,harley,lisitan,fogli}.

If one chooses to ignore the possibility of sterile neutrinos, then
one should use in eq..~(\ref{eqn:chidiff}) $\Delta\chi^2(\mbox{C.L.,
2~d.o.f.})$ instead of $\Delta\chi^2(\mbox{C.L., 3~d.o.f.})$.
The corresponding numbers for 2 d.o.f are 
$\Delta\chi^2(\mbox{C.L., 2~d.o.f.}) = 4.61$, $5.99$, $9.21$, and
$11.83$ for C.L.~= 90\%, 95\%, 99\% and 99.73\% ($3\sigma$), For the
case in which only active neutrinos are considered, the numbers for
$\chi^2_{\rm min}$ given in the various tables in this paper can be
used together with the above values of $\Delta\chi^2(\mbox{C.L.,
2~d.o.f.})$ to determine the smaller allowed regions.

\section{Global Solutions including Rates and Day-Night Spectra}
\label{sec:globalplus}

We describe in this section global solutions obtained by using all the
relevant  solar neutrino and reactor data.

\FIGURE[!ht]{
\centerline{\psfig{figure=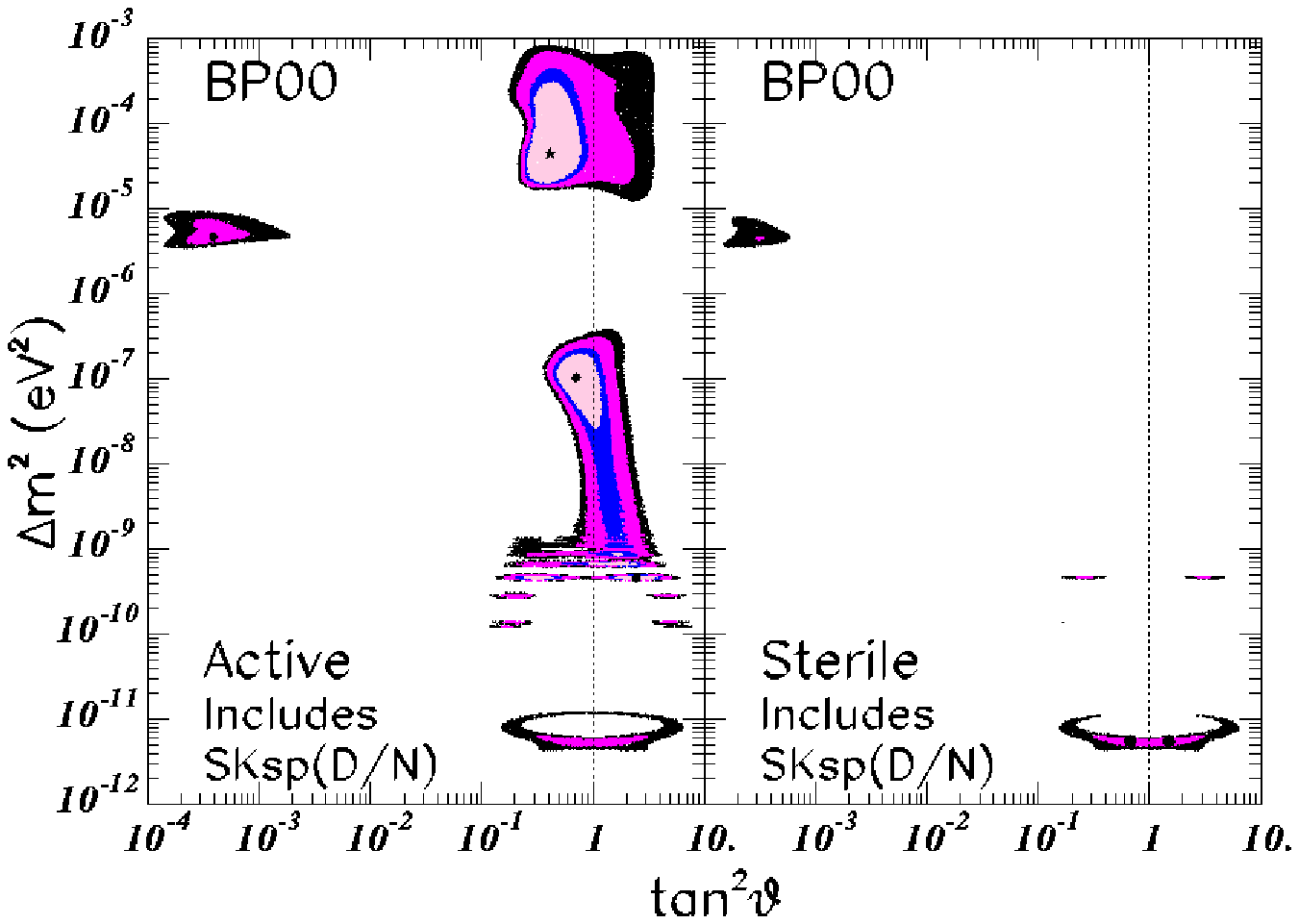,width=4.5in}}
\caption{{\bf Global solutions including all available solar neutrino
data.} The input data include the total rates from the
SNO~\cite{sno2001}, Chlorine~\cite{chlorine}, and Gallium
(averaged)~\cite{gallex,gno,sage} experiments, as well as the recoil
electron energy spectrum measured by Super-Kamiokande~\cite{superk}
during the day and separately the energy spectrum measured at
night. The C.L. contours shown in the figure are $90$\%, $95$\%,
$99$\%, and $99.73$\% ($3\sigma$). The allowed regions are limited by
the Chooz reactor measurements~\cite{chooz} to mass values below
$\sim$ (7--8) $\times 10^{-4} {\rm eV^2}$. The local best-fit points
are marked by dark circles. The theoretical errors for the BP00
neutrino fluxes are included in the analysis.
\label{fig:global}} }

Figure~\ref{fig:global} shows the globally allowed solutions when the
Super-Kamiokande recoil energy spectrum during the day and,
separately, the energy spectrum at night are included in addition to
the total rates in the Chlorine~\cite{chlorine}, Gallium
(averaged)~\cite{gallex,gno,sage}, and SNO~\cite{sno2001}
experiments. In order to avoid double counting the Super-Kamiokande
total rate(cf. ref.~\cite{bks2001} for a discussion of this point), we
have not included the Super-Kamiokande rate~\cite{superk} in addition
to the Super-Kamiokande day and night energy spectra, which each
contain their own absolute normalizations. We have used the Chooz
reactor bound~\cite{chooz} to limit the allowed solutions to  mass
values below  $\sim$ (7--8) $\times 10^{-4} {\rm eV^2}$.

Figure~\ref{fig:global} is our currently preferred global solution.

All eight of the allowed solutions for active and sterile neutrinos
that existed before the SNO CC measurement (e.g., ref.~\cite{bks2001})
are still allowed at $3\sigma$ after including the results of the SNO
CC measurement.

\TABLE{
\centering
\caption{\label{tab:globalbestbp00} {\bf Best-fit global oscillation
parameters with all solar neutrino data.}  This table corresponds to
the global solution illustrated in figure~\ref{fig:global}.  The
differences of the squared masses are given in ${\rm eV^2}$.  The
number of degrees of freedom is 38 [38(spectrum) + 3(rates)
$-$3(parameters: $\Delta {\rm m}^2$, $\theta$, active--sterile
admixture)].  The goodness-of-fit given in the last column is
calculated using the value of $\chi^2/{\rm d.o.f}$ for each allowed
solution.  The BP00 best-fit fluxes and their estimated errors have
been included in the analysis. The rates from the GALLEX/GNO and SAGE
experiments have been averaged to provide a unique data point.
The goodness-of-fit given in the
last column is calculated from the value of $\chi^2/{\rm d.o.f}$ at
each local minimum (i. e., for LMA, SMA, VAC, LOW, etc.).
}
\begin{tabular}{lcccc} 
\noalign{\bigskip}
\hline 
\noalign{\smallskip}
Solution&$\Delta m^2$&$\tan^2(\theta)$& $\chi^2_{\rm min}$ &g.o.f. \\
\noalign{\smallskip}
\hline
\noalign{\smallskip}
LMA& $4.5\times10^{-5}$  &$4.1\times10^{-1}$ & 35.3 &$59$\% \\
LOW& $1.0\times10^{-7}$  &$7.1\times10^{-1}$ & 38.4 &$45$\%\\ 
VAC& $4.6\times10^{-10}$ &$2.4\times10^{0}$ & 39.0 &$42$\%\\ 
SMA& $4.7\times10^{-6}$  &$3.9\times10^{-4}$ & 45.4 &$19$\%\\
Just So$^2$ & $5.5\times10^{-12}$  
&\hskip -6pt$0.67 (1.5) \times10^{0} $ & 45.7 &$18$\%\\
 Sterile Just So$^2$ & $5.5\times10^{-12}$  &\hskip -6pt$0.67 (1.5) \times10^{0} $ & 
45.8 &$18$\%\\
 Sterile SMA & $4.5\times10^{-6}$ & $3.1\times10^{-4}$ & 46.6 & $16$\%\\
 Sterile VAC & $4.7\times10^{-10}$ & $2.7\times10^{-1}$ & 47.2 & $15$\%\\  
\noalign{\smallskip}
\hline
\end{tabular}
}

\FIGURE[!ht]{
\centerline{\psfig{figure=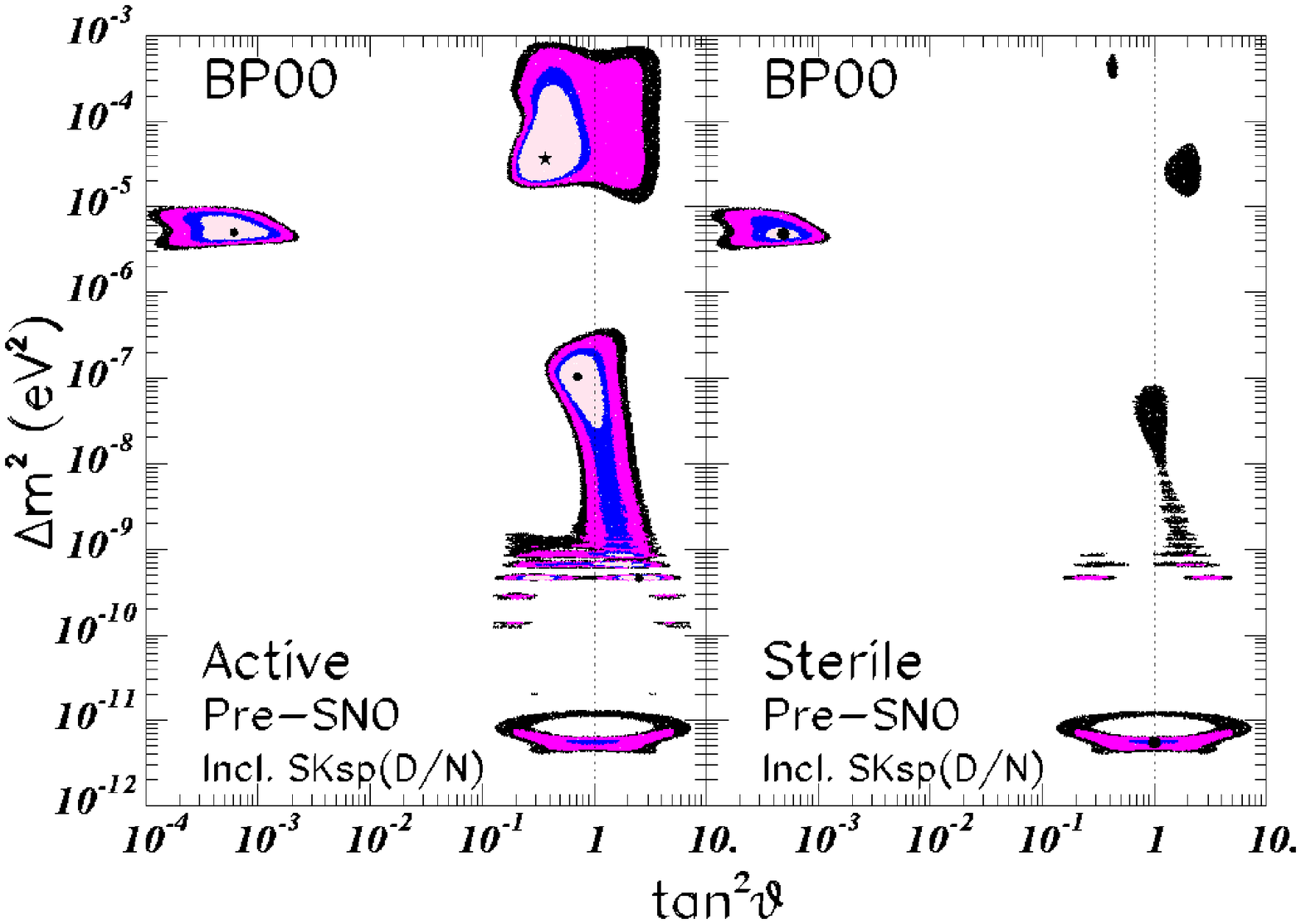,width=4.5in}}
\caption{{\bf Pre-SNO global solutions.} The input data and the analysis
procedures are the same as used in producing
figure~\ref{fig:global}, except that we have not included the SNO CC
measurement in constructing figure~\ref{fig:presno}.
\label{fig:presno}} }

\FIGURE[!ht]{
\centerline{\psfig{figure=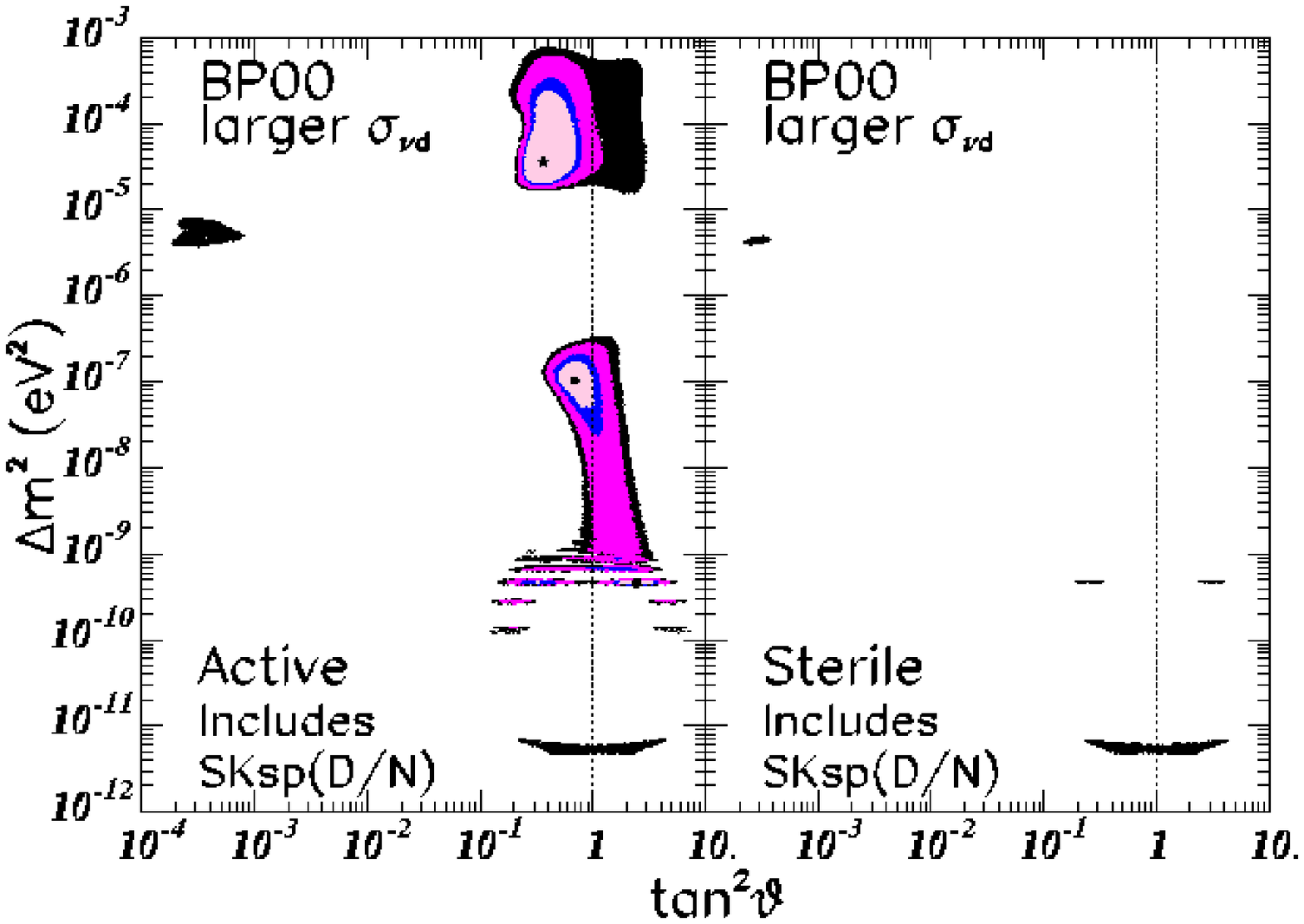,width=4.5in}}
\caption{{\bf Global solutions including all available solar neutrino
data, with enhanced CC cross section for deuterium.}
The input date are the same as in figure~\ref{fig:global} except that
we have used a $4$\% large CC cross section for neutrino absorption
on deuterium. 
\label{fig:globaln}} }

Table~\ref{tab:globalbestbp00} gives the best-fit parameters for each
of the eight allowed oscillation solutions illustrated in
figure~\ref{fig:global}.  We give in table~\ref{tab:globalbestbp00}
the values of $\Delta m^2$, $\tan^2\theta$, $\chi^2_{\rm min}$, and the
goodness-of-fit for each of the best-fit points. The solutions
described by the acronyms like LMA and Just So$^2$ can be identified
in figure~\ref{fig:global} with the aid of
table~\ref{tab:globalbestbp00} by  using the association between the
solution acronyms (column 1) and the values of $\Delta m^2$ (column 2).

The LMA active solution is the best-fit, but is only slightly better
than the LOW and VAC active solutions. The SMA solution is a
significantly less good fit than the LMA and LOW solutions. The
goodness-of-fit ranges from $59$\% for the LMA active solution to
$15$\% for the Sterile VAC solution, all satisfactory fits to the
available data.

The five active solutions, LMA, SMA, LOW, VAC, and Just So$^2$ all
appear clearly in the left hand panel of figure~\ref{fig:global}, although
the allowed area of the SMA solution has shrunk significantly relative
to the pre-SNO situation (cf.
refs.~\cite{bks2001,cc2001}).\footnote{We caution the reader that the
comparison with previous results is not entirely straightforward since
we have included in the $\chi^2$ analysis an additional parameter
($\cos \eta
= 1,0$, for active or sterile neutrinos) and we measure all
departures from $\chi^2_{\rm min}$, the best-fit value with either
active or sterile neutrinos.  We have also made a few minor
adjustments in the analysis code.} The largest reduction in allowed
area occurs for the Sterile SMA, which is barely visible at $3\sigma$
C.L. in the right hand panel of figure~\ref{fig:global}. The other
sterile solutions are not much affected at $3\sigma$ by the SNO CC
measurement, although they have also become somewhat less likely.
The Just So$^2$ solution for both active and sterile 
oscilaltions has also worsened after the
inclusion of the SNO CC rate.

For oscillations among only active neutrinos, the statistical criteria
used to define the allowed regions are not unique.  One could assume
that there are no sterile neutrinos (one cannot assume that are no
active neutrinos.).  If one knows {\it a priori} that Nature does not
contain light sterile neutrinos, then we live in the slice of solution
space corresponding to $\cos^2\eta = 1$.  Given this {\it a priori}
knowledge, the regions at a given C.L. would be 2-dimensional and then
the analysis for the active neutrino oscillations would be the
standard one with regions at a given C.L. defined in terms of
$\Delta\chi^2(\mbox{C.L., 2~d.o.f.}$) (see discussion in
section~\ref{sec:calculational} following eq.~\ref{eqn:chidiff}). In this
approach, the active SMA region would appear only at $3\sigma$ since
$\chi^2_{\rm SMA}-\chi^2_{\rm LMA}=10.1$.
  
We discuss explicitly in section~\ref{sec:comparisons} (see especially
figure~\ref{fig:twodof}) the allowed solutions in the case in which
one knows, by some other means, that sterile neutrinos do not exist.

How much has the measurement by SNO of the CC rate affected the
globally allowed solutions?  Figure~\ref{fig:global} and
figure~\ref{fig:presno}, when compared, answer this question.

The SNO measurement has not changed qualitatively the allowed solution
space for active neutrinos. Nevertheless, the SMA solution is now less
likely since it is difficult for SMA to fit simultaneously the
Super-Kamiokande flat spectrum and the SNO CC measurement.  The
comparison between the SNO CC measurement and the Super-Kamiokande
total rate indicates the presence of some NC contribution to the
measured Super-Kamiokande rate.  For the SMA solution, the neutral
current contribution is only predicted to be large enough for the
larger mixing angles.  For the larger SMA mixing angles, the survival
probability of $^8$B neutrinos rises with energy (for this reason the
SMA region in figure~\ref{fig:ratesonly} for the rates-only analysis
is shifted towards larger mixing angles). However, this predicted
growth of the survival probability with energy conflicts with the
Super-Kamiokande measurement~\cite{superk} that is consistent with an
undistorted recoil energy spectrum.

The evidence for a NC contribution in the Super-Kamiokande rate also
disfavors oscillations into sterile neutrinos.  The LMA and LOW
sterile neutrino solutions, previously allowed at $3\sigma$ are now
disfavored at this C.L. and the allowed region of the SMA sterile
solution has been reduced in size. The fit is also worsened for the
Just So$^2$ solution (from $\chi^2_{\rm Just\,\, So^2}-\chi^2_{\rm
LMA}=7.6$ in the pre-SNO analysis to $\chi^2_{\rm Just \,\,
So^2}-\chi^2_{\rm LMA}=10.4$ in the post-SNO analysis). We will return
to this point in Sec.~\ref{subsec:delicate}.

\TABLE[!h]{
\caption{\label{tab:globalbestnbp00} {\bf Best-fit global oscillation
parameters with an enhanced CC cross section for deuterium,
corresponding to figure~\ref{fig:globaln}.} The input data used in
constructing this table were the same as were used in constructing
table~\ref{tab:globalbestbp00} except that for this table we adopted a
$4$\% larger CC cross section for deuterium. }
\begin{tabular}{lcccc} 
\noalign{\bigskip}
\hline 
\noalign{\smallskip}
Solution&$\Delta m^2$&$\tan^2(\theta)$& $\chi^2_{\rm min}$ &g.o.f. \\
\noalign{\smallskip}
\hline
\noalign{\smallskip}
LMA& $3.7\times10^{-5}$  &$3.7\times10^{-1}$ & 34.7 &$62$\% \\
LOW& $1.0\times10^{-7}$  &$6.9\times10^{-1}$ & 39.2 &$42$\%\\ 
VAC& $4.6\times10^{-10}$ &$2.4\times10^{0}$ & 39.7 &$39$\%\\ 
SMA& $4.6\times10^{-6}$  &$3.4\times10^{-4}$ & 47.0 &$15$\%\\
 Just So$^2$ & $5.5\times10^{-12}$  &\hskip -6pt$0.67 (1.5) \times10^{0} $ & 46.8 &$15$\%\\
 Sterile Just So$^2$ & $5.5\times10^{-12}$  &\hskip -6pt$0.67 (1.5) \times10^{0} $ & 47.0 &$15$\%\\
 Sterile SMA & $4.5\times10^{-6}$ & $2.9\times10^{-4}$ & 48.2 & $12$\%\\
 Sterile VAC & $4.7\times10^{-10}$ & $2.7\times10^{-1}$ & 48.5 & $12$\%\\  
\noalign{\smallskip}
\hline
\end{tabular}
}

What is the effect of the possibly enhanced cross section discussed in
ref.~\cite{beacom} and in section~\ref{sec:calculational}?
Figure~\ref{fig:globaln} and table~\ref{tab:globalbestnbp00} show the
results that are obtained when a CC cross section enhanced by $4$\% is
assumed for deuterium. It is instructive to compare directly
figure~\ref{fig:globaln} and table~\ref{tab:globalbestnbp00} with
figure~\ref{fig:global} and table~\ref{tab:globalbestbp00}. Doing so,
we see that no qualitative changes are induced by using the larger CC
cross section for deuterium. The main quantitative change is that the allowed
regions for the SMA solutions, both active and sterile, become
slightly smaller and less statistically likely when the larger CC
cross section is assumed. All of the sterile neutrino solutions become
slightly less good fits.

The wonderful measurement of the $\nu_e$ flux of $^8$B neutrinos that
has recently been reported by SNO~\cite{sno2001} is the first
quantitative result reported by this collaboration.  It is therefore
plausible that the error on the $\nu_e$ flux will decrease with
time as the systematic uncertainties become better understood and the
statistical errors are reduced by counting more events.  In an
uncontrolled burst of optimism, we have hypothesized that the quoted
experimental error on the $\nu_e$  flux will be ultimately
reduced by a factor of three while the best-estimate value for the
$\nu_e$ flux will be unchanged.

Figure~\ref{fig:ccdividedby3} shows the effect on the globally allowed
solutions of reducing the total error on the CC flux measurement of
SNO by a factor of three relative to the total error quoted in
ref.~\cite{sno2001}.  Comparing figure~\ref{fig:ccdividedby3} with
figure~\ref{fig:global}, we see that a factor of three improvement in
the quoted error could eliminate at $3\sigma$ both the active and the
sterile SMA solutions and disfavor (rule out at $3\sigma$) the active
(sterile) Just So$^2$ solution. The allowed LMA and LOW solutions
would not be much affected by even a factor of three reduction in the
error in the $\nu_e$ flux, the main effect being a modest reduction of
the area of the allowed LMA solution in the $\Delta
m^2$-$\tan^2\theta$ plane.

\FIGURE[!t]{
\centerline{\psfig{figure=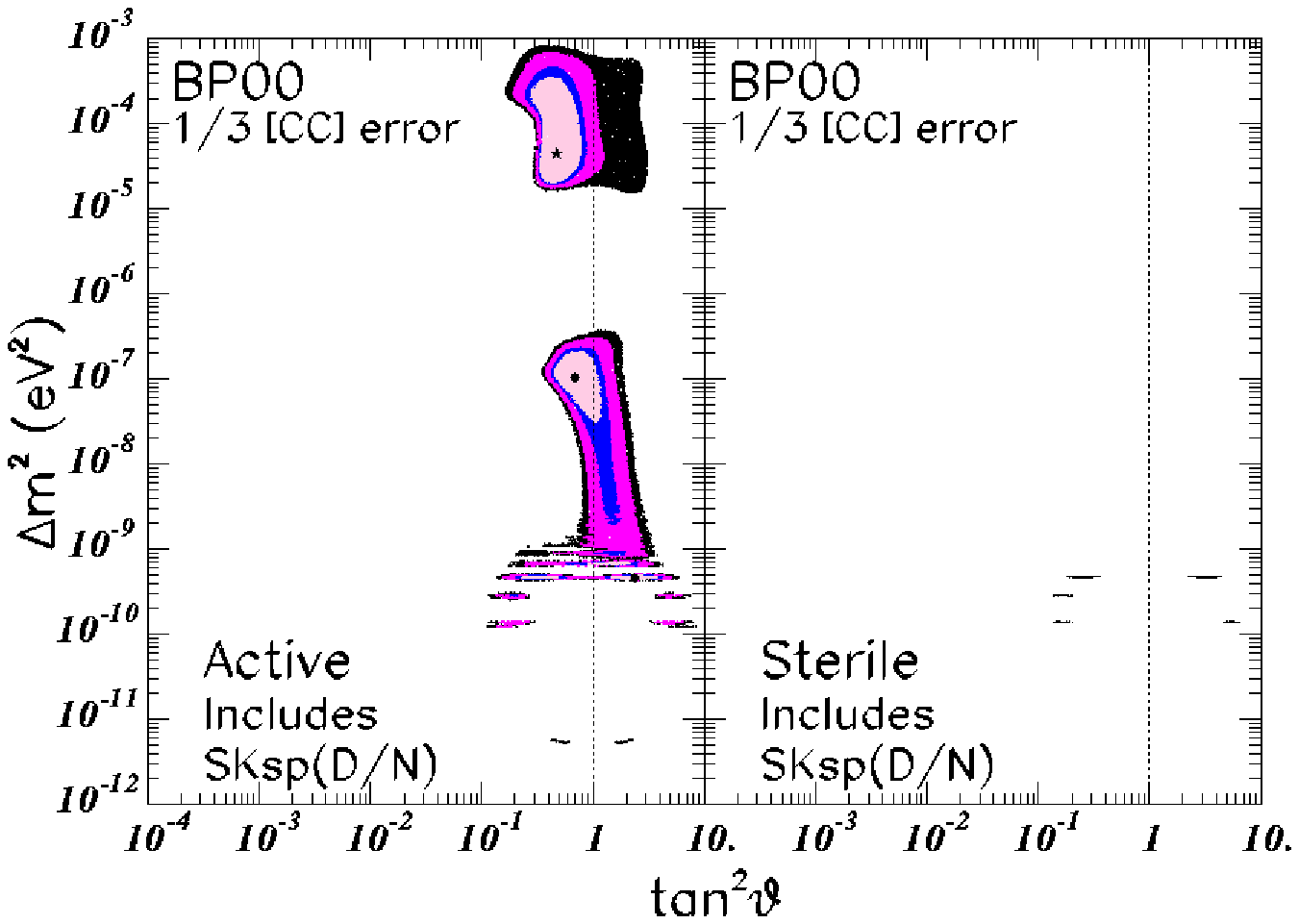,width=4.5in}}
\caption{{\bf Global solutions with error on SNO CC flux reduced by a
factor of three.} The input data are the same as for
figure~\ref{fig:global} except that the error on the $\nu_e$ flux of
$^8$B neutrinos for SNO was artificially reduced by a factor of
three.\label{fig:ccdividedby3}} }

Does it make a difference if we constrain the $^8$B neutrino flux
according to the predictions of the standard solar model? This is an
important question to answer, since one can give reasonable arguments
on both sides of the question as to whether or not it is more
appropriate to constrain the $^8$B neutrino flux. Arguing in favor,
one can point out that the standard solar model provides relevant
information about the solar interior and about the rates of solar
fusion reactions. Moreover, the standard solar model is remarkably
successful in predicting results in agreement with helioseismological
measurements~\cite{bp2000}. On the other hand, we would like to
determine the $^8$B neutrino flux independent of solar model
considerations. For a model-independent analysis, we must allow the
$^8$B neutrino flux to vary freely.

Figure~\ref{fig:globalfree} shows the result of an analysis that is
identical to our standard analysis (cf. figure~\ref{fig:global})
except that in constructing figure~\ref{fig:globalfree} the $^8$B
neutrino flux was not constrained by solar model predictions. We have
checked that the global minimum in this case lies in the LMA region
for purely active neutrinos, as is also the case for the analysis with
the constrained BP00 $^8$B neutrino flux.  For the rates-only analysis
discussed in the next section, the $\chi^2$ minimum for the
unconstrained $^8$B neutrino flux analysis has non-vanishing
components of both active and sterile neutrinos, in agreement with the
results of ref.~\cite{barger2001} (see also the discussion in
Sec.~\ref{sec:comparisons}).

\FIGURE[!h]{
\centerline{\psfig{figure=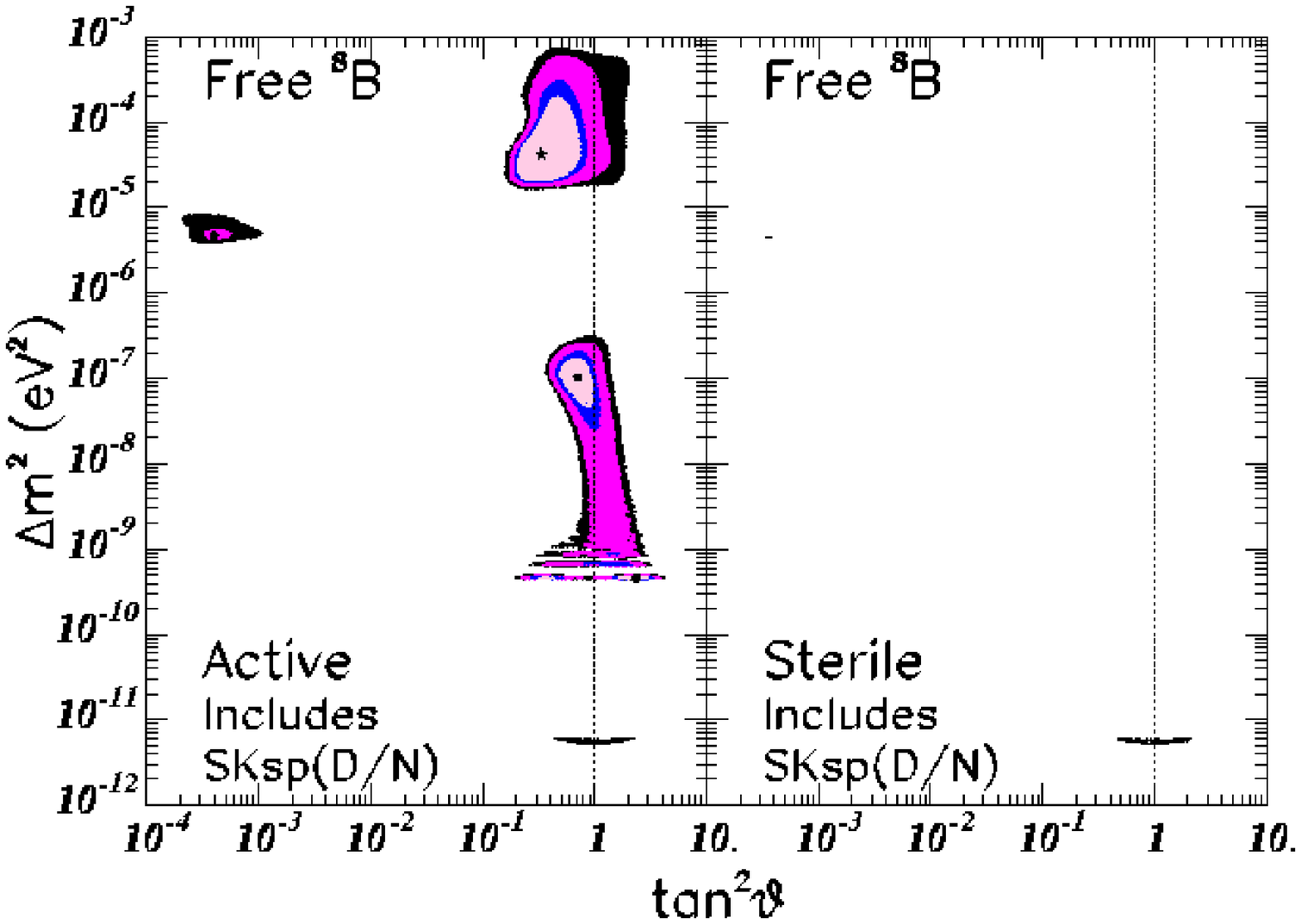,width=4.5in}}
\caption{{\bf Global solutions, with unconstrained$^8$B neutrino flux,
including all available solar neutrino
data.} The input data and the analysis are the same as were used in
constructing figure~\ref{fig:global} except that for the present
figure no constraint was placed on the $^8$B neutrino flux.
\label{fig:globalfree}} } 

The three most favored solutions in figure~\ref{fig:globalfree} all
involve large mixing angles; they are (with their associated g.o.f):
LMA ($59$\%), LOW ($39$\%), and VAC ($35$\%).  The g.o.f. for all the
other solutions is less than $20$\%.

Comparing figure~\ref{fig:globalfree} and figure~\ref{fig:global}, we
see that the favored, large angle solutions, LMA and LOW, are not
changed significantly. However, the allowed regions for the SMA and
Just So$^2$ solutions, both active and sterile, are reduced in size by
performing a $^8$B-free analysis. The sterile SMA allowed region is
reduced to almost a point in figure~\ref{fig:globalfree}. To find the
sterile SMA in figure~\ref{fig:globalfree}, one has to know where to
look, namely, near the $\Delta m^2$ and $tan^2 \theta$ for which the
best-fit SMA solution appears for active neutrinos. Also, the small
sterile VAC region that appears in figure~\ref{fig:global} disappears
entirely in figure~\ref{fig:globalfree}.

It is difficult to anticipate intuitively the quantitative difference
between global solutions obtained without and without constraints on
the $^8$B neutrino flux. This is particularly true for the less robust
solutions, like all of the sterile solutions and the Just So$^2$
active solution. Two factors work in opposite directions in
determining $\chi^2$: 1) the free normalization in the $^8$B neutrino
flux; and 2) the removal of the theoretical error in the flux. The
second effect seems to be generally more important.

\FIGURE[!t]{
\centerline{\psfig{figure=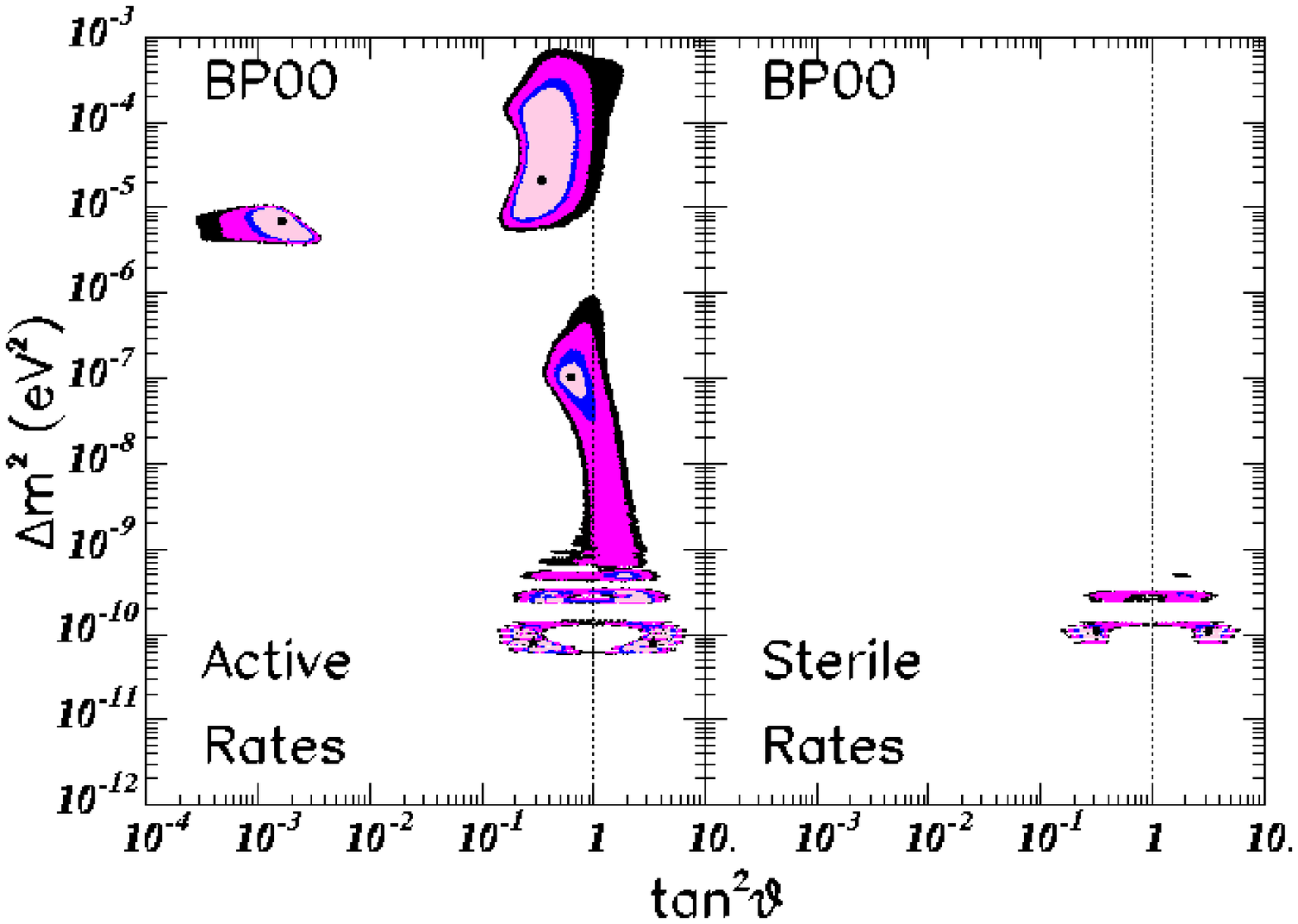,width=4.5in}}
\caption{{\bf Global solutions for the rates only.} The input data are
the total rates measured in the SNO CC~\cite{sno2001},
Chlorine~\cite{chlorine}, SAGE~\cite{sage} + (GALLEX~\cite{gallex} +
GNO~\cite{gno}), and Super-Kamiokande~\cite{superk} experiments.  The
C.L. contours shown in the figure are $90$\%, $95$\%, $99$\%, and $99.73$\%
($3\sigma$). The local best-fit points are marked by dark circles. The
theoretical errors for the BP00 neutrino fluxes are included in the
analysis.\label{fig:ratesonly}} }

\section{Rates-only Global Solutions}
\label{sec:ratesonly}

We describe in this section global solutions obtained by considering
only the total rates in the Chlorine, GALLEX/GNO, SAGE, Super-Kamiokande,
and SNO (CC) solar neutrino experiments. We also evaluate the effects
of varying the $^8$B neutrino flux unconstrained by solar model
predictions and of increasing the CC cross section for SNO.

Figure~\ref{fig:ratesonly} and table~\ref{tab:ratbp00} present the
allowed solution regions and the best-fit parameters for the global
analysis of only the total rates. The best-fit point is in the VAC
region with the LMA solution giving a rather similar $\chi^2_{\rm
min}$. 

\TABLE{\centering
\caption{\label{tab:ratbp00} {\bf Best-fit parameters for total rates
only, corresponding to figure~\ref{fig:ratesonly}.}  The format of
this table is the same as for table~\ref{tab:globalbestbp00}.  The two
rates measured by the Gallium experiments, (GALLEX~\cite{gallex} and
GNO~\cite{gno}) and SAGE~\cite{sage}, are included separately.  The
number of degrees of freedom is 2 [5(rates) $-$3(parameters: $\Delta
{\rm m}^2$, $\theta$, $O = $ active or sterile neutrino)].  The
Sterile SMA solution does not appear in figure~\ref{fig:ratesonly}
because $\chi^2_{\rm min}$ is too large for this solution, but the
result is given in the table for general information. }
\begin{tabular}{lcccc} 
\noalign{\bigskip}
\hline 
\noalign{\smallskip}
Solution&$\Delta m^2$&$\tan^2(\theta)$& $\chi^2_{\rm min}$ &g.o.f. \\
\noalign{\smallskip}
\hline
\noalign{\smallskip}
VAC& $7.9\times10^{-11}$  &$0.29 (3.45)$ & 3.50 &$17$\% \\
LMA& $2.1\times10^{-5}$  &$3.4\times10^{-1}$ & 3.99 &$14$\% \\
SMA& $6.9\times10^{-6}$  &$1.6\times10^{-3}$ & 5.25 &$7$\%\\
LOW& $9.7\times10^{-8}$  &$6.5\times10^{-1}$ & 8.61 &$1.4$\%\\ 
Sterile VAC & $1.1\times10^{-10}$ & $0.29 (3.45)$ & 10.1 & $0.63$\%\\  
Sterile SMA & $4.9\times10^{-6}$ & $5.3\times10^{-4}$ & 18.0 & $0.01$\%\\
\noalign{\smallskip}
\hline
\end{tabular}
}

For the LMA, SMA, and LOW solutions, the results appear superficially
similar to what is obtained when the Super-Kamiokande spectral data for
both the day and the night are included in the analysis
(cf. figure~\ref{fig:global} and figure~\ref{fig:ratesonly}).
However, the g.o.f. is poorer (cf. table~\ref{tab:globalbestbp00} and
table~\ref{tab:ratbp00}) when the spectral data are included, which
reflects the fact that solutions which predict a flat spectrum in the
energy region studied by Super-Kamiokande are favored.

\FIGURE[!ht]{
\centerline{\psfig{figure=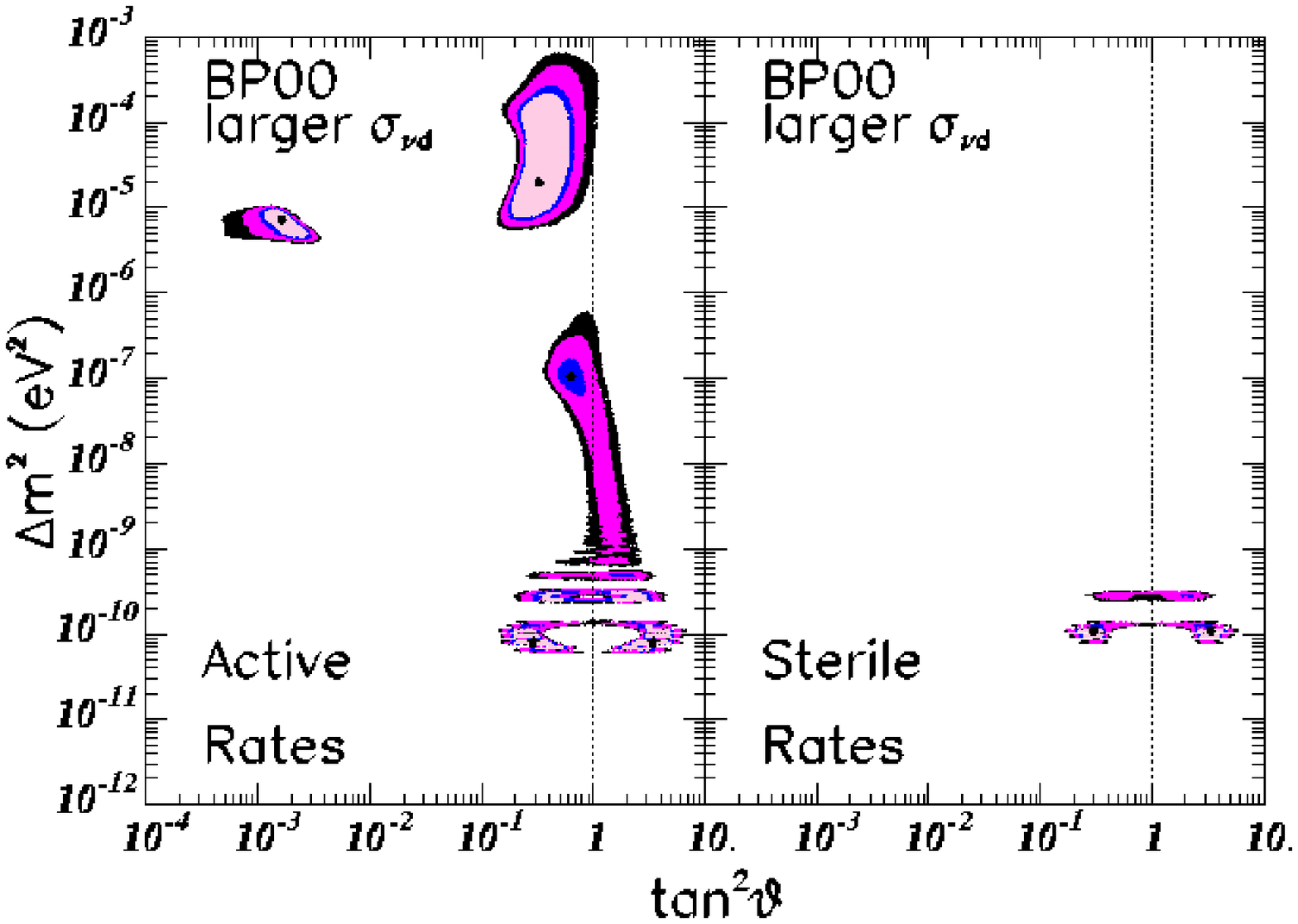,width=4.5in}}
\caption{{\bf Global solutions for the rates only, with an
enhanced CC cross section for deuterium.} The input data and the
analysis procedures are the same as used in producing
figure~\ref{fig:ratesonly}, except that we have used for the present
figure a $4$\% larger CC cross section for neutrino absorption by
deuterium. \label{fig:ratesnonly}} }

The principal changes that result from omitting the spectral and the
day-night data are for the vacuum solutions near $10^{-10} {\rm
~eV^2}$, the Just So$^2$ solution, and the Sterile SMA solution.  The
vacuum solutions are much more prominent when the spectral data are
not considered. This is a well-know effect and has been noted in
many previous analyses. The Just So$^2$ solution is missing when only
the rates are considered, which is largely due to the
imposition of the BP00 theoretical constraint on the $^8$B
flux~\cite{bks2001}. The sterile SMA solution also disappears
completely at $3\sigma$ (just barely, see table~\ref{tab:ratbp00}).

Figure~\ref{fig:ratesnonly} and table~\ref{tab:ratnbp00} show that
enhancing the CC cross section by $4$\% has very little effect on
the allowed regions when only the data on total event rates is
considered. This result is seen most clearly by comparing directly
figure~\ref{fig:ratesonly} and table~\ref{tab:ratbp00} with
figure~\ref{fig:ratesnonly} and table~\ref{tab:ratnbp00}. 

\TABLE[]{
\caption{\label{tab:ratnbp00} {\bf Best-fit parameters for total rates
only with enhanced CC cross section for deuterium, corresponding to
figure~\ref{fig:ratesnonly}.}  The input data for this analysis are
the same as for the analysis described by table~\ref{tab:ratnbp00}
except that a $4$\% larger CC cross section for deuterium was used
here.}
\begin{tabular}{lcccc} 
\noalign{\bigskip}
\hline 
\noalign{\smallskip}
Solution&$\Delta m^2$&$\tan^2(\theta)$& $\chi^2_{\rm min}$ &g.o.f. \\
\noalign{\smallskip}
\hline
\noalign{\smallskip}
VAC& $7.9\times10^{-11}$  &$0.29 (3.45)$ & 3.07 &$21$\% \\
LMA& $2.0\times10^{-5}$  &$3.3\times10^{-1}$ & 3.45 &$18$\% \\
SMA& $7.3\times10^{-6}$  &$1.6\times10^{-3}$ & 7.09 &$2.9$\%\\
Sterile VAC & $1.1\times 10^{-10}$ & $0.29 (3.45)$ & 9.12 & $1.1$\%\\  
LOW& $1.1\times10^{-7}$  &$6.3\times10^{-1}$ & 9.45 &$0.89$\%\\ 
Sterile SMA & $4.7\times 10^{-6}$ & $4.6\times10^{-4}$ & 20.7 & $0.003$\%\\
\noalign{\smallskip}
\hline
\end{tabular}
}

\FIGURE[!ht]{
\centerline{\psfig{figure=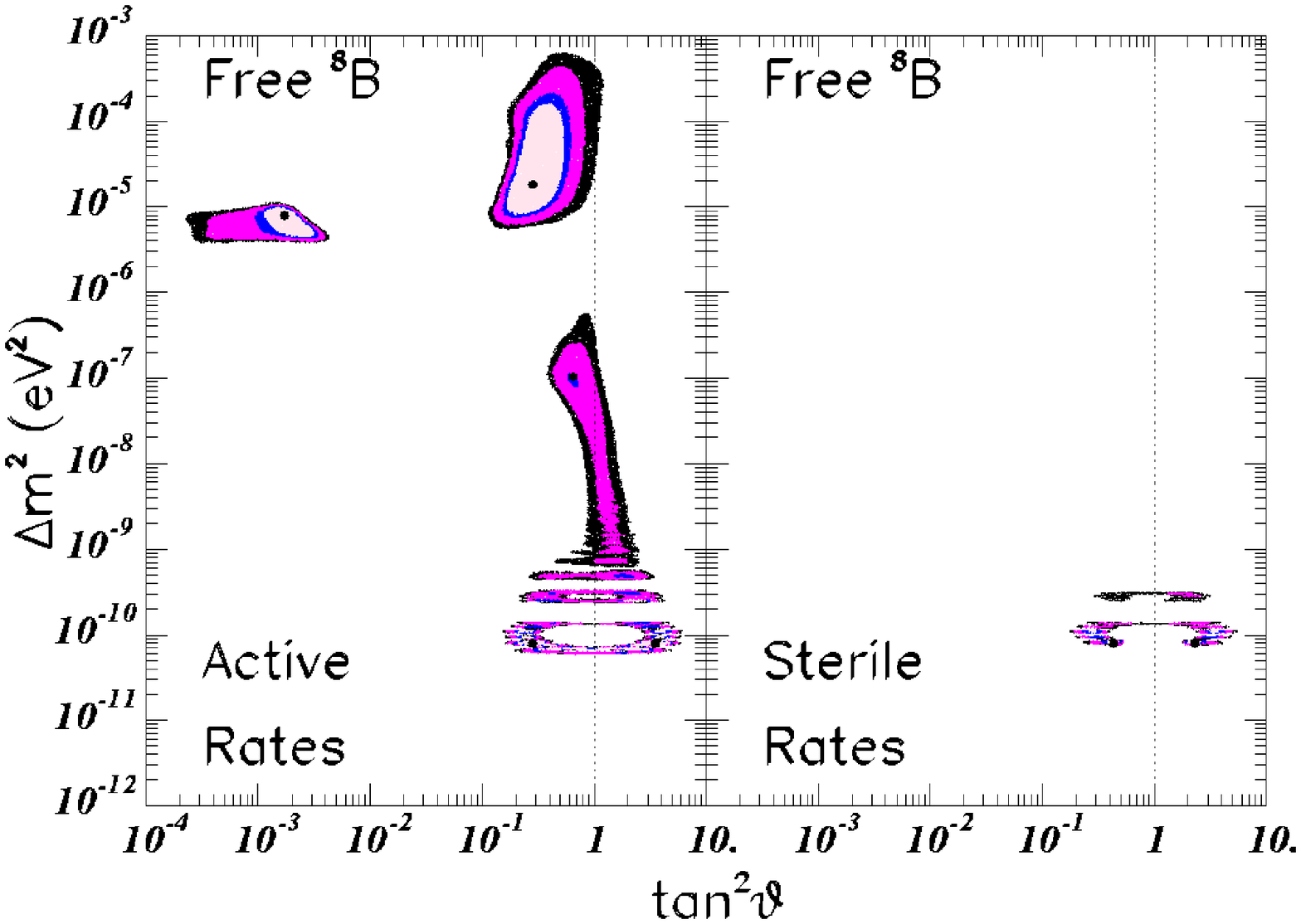,width=4.5in}}
\caption{{\bf Global solutions for the rates only, with an
unconstrained $^8$B neutrino flux.} The input data and the analysis
procedures are the same as used in producing
figure~\ref{fig:ratesonly}, except that the $^8$B flux is allowed to
vary without considering the predicted flux or errors of the standard
solar model. \label{fig:ratb8free}} }

Figure~\ref{fig:ratb8free} shows the allowed solutions that are
obtained when the $^8$B flux is permitted to vary freely, without
considering the constraints implied by the standard solar model.
There are no major differences between the allowed regions shown in
figure~\ref{fig:ratb8free} and the allowed regions shown in
figure~\ref{fig:ratesonly}, for which the $^8$B neutrino flux was
constrained by the BP00 flux predictions.

The minimum $\chi^2$ for the rates-only, free $^8$B case shown in
figure~\ref{fig:ratb8free} lies in the LMA region, but for a finite
value of the active-sterile admixture. The sterile component increases
as $f_{\rm B}$ increases, as described in ref.~\cite{barger2001}.  We
have restricted ourselves to normalization factors $f_{\rm B}\leq 2$,
because (cf. section~\ref{subsec:defnchi2}) larger values of $f_{\rm
B}\leq 2$ are ruled out at a $3\sigma $ C.L. by the Super-Kamiokande
day-night recoil energy spectrum and are implausible from an
astrophysical point of view.  With this restriction, the global
minimum is found at $\sin^2\eta=0.55$, with $\chi^2_{\rm min}=0.79$,
$\Delta m^2=1.3 \times 10^{-5}{\rm \, eV^2}$, $tan^2 \theta = 0.17$,
and $f_B = 1.9$. The regions shown in figure~\ref{fig:ratb8free} are
defined with respect to this global minimum.  We have verified that if
even larger $f_{\rm B}$ were allowed the minimum would occur at larger
$\sin^2\eta$, in agreement with the results of ref.~\cite{barger2001}.

We caution, however, that for figure~\ref{fig:ratb8free} the LMA
best-fit point lies in the lower part of the LMA region ($\Delta
m^2=1.8 \times 10^{-5}{\rm \, eV^2}$ for pure active neutrinos,
$\sin^2\eta=0$, moving down to $\Delta m^2=1.2 \times 10^{-5}{\rm \,
eV^2}$ for $\cos^2\eta=0.55$).  This part of the LMA region is
disfavored at $3\sigma$ by the Super-Kamiokande day-night data. For
those values of the parameters, the Earth regeneration effects are
important and the day-night asymmetry is much larger than observed.
Thus we conclude that the results of ref.~\cite{barger2001} is correct
only the total rates are considered, but cannot be extended to the
case in the observed day-night spectral energy data are included.  We
will return to this point in Sec.~\ref{subsec:mixtures}.

Finally, we have investigated the effect of reducing the total error
on the $\nu_e$ flux measured by SNO. We have supposed, in the same
burst of wild optimism that prevailed previously (see
section~\ref{sec:globalplus}, especially the discussion of
figure~\ref{fig:ccdividedby3}), that the total error on the $\nu_e$
flux is reduced by a factor of three while the best-estimate for
the flux remains unchanged from the value quoted by
ref.~\cite{sno2001}. We find results consistent with the discussion of 
figure~\ref{fig:ccdividedby3}, which applies to the case when all
the available solar neutrino data are included in the analysis. In the
present case, we find that the SMA solution is greatly reduced in
area. The LMA and LOW solutions are not qualitatively affected.

\section{Comparisons and Clarifications}
\label{sec:comparisons}

In this paper, we have done several things that are relatively new in the
context of two-component solar neutrino oscillations: *) include the
initial SNO CC data~\cite{sno2001}; *) implement a
prescription first used in ref.~\cite{bks2001} for handling
consistently the rate normalization and the spectral data for the
Super-Kamiokande experiment; and *) treat active and sterile neutrino
oscillations on an equal basis~\cite{four,harley,lisitan,fogli}.
We have already illustrated the effect of the SNO CC measurement by
comparing figure~\ref{fig:global} with figure~\ref{fig:presno}. 

We will clarify in this section the role of the later two aspects of
the analysis, the consistent treatment of the rate normalization and
the spectral energy distribution, and the level playing field
treatment of active and sterile neutrinos.  We also compare our
results with two recent and very interesting papers, by Barger,
Marfatia, and Whisnant~\cite{barger2001} and by Fogli, Lisi,
Montanino, and Palazzo~\cite{foglipostsno}.  In all cases, we
reproduce the results of ref.~\cite{barger2001} and
ref.~\cite{foglipostsno} by analyzing the data in the manner described
in these references. The present section was not included in the
original version of our paper submitted to the journal, because our
analysis was essentially simultaneous with the work of Barger et
al. and with Fogli et al.

We begin by discussing in section~\ref{subsec:nosterile} the effect of
rejecting {\it a priori} the existence of sterile neutrinos and then
discuss in section~\ref{subsec:delicate} the effect of different
treatments of the measured normalization of the $^8$B neutrino flux
and the associated influence of the standard solar model constraint on
the $^8$B neutrino flux. The discussions in
section~\ref{subsec:nosterile} and section~\ref{subsec:delicate} are
most relevant to the analysis in ref.~\cite{foglipostsno}. In
section~\ref{subsec:mixtures} we show that the small day-night
difference measured by Super-Kamiokande disfavors at $3\sigma$ the
primarily sterile neutrino mixtures that were found in
ref.~\cite{barger2001} when analyzing only the total event rates.
  
\subsection{{\it A priori} rejecting sterile neutrinos}
\label{subsec:nosterile}

Assuming that only active neutrinos exist, 
figure~\ref{fig:twodof} shows the global solutions with and without
constraining the $^8$B flux by the standard solar model~\cite{bp2000}.
The two panels in figure~\ref{fig:twodof} should be compared,
respectively, with figure~\ref{fig:global} and with
figure~\ref{fig:globalfree} and with figure~6 of ref.~\cite{foglipostsno}.

The main effect of rejecting sterile neutrinos {\it a priori} is to
reduce the area of the allowed regions; the reduction is particularly
apparent for the less robust SMA and Just So$^2$ solutions.  For the
case in which the $^8$B neutrinos are constrained by the SSM, both
regions become only marginally allowed at $3\sigma$.
For the unconstrained $^8$B neutrinos, only the
SMA marginally survives at 3$\sigma$ while the Just So$^2$ solution is
no longer allowed. This trend is consistent with the results reported
in ref.~\cite{foglipostsno}.

\FIGURE[!ht]{
\centerline{\psfig{figure=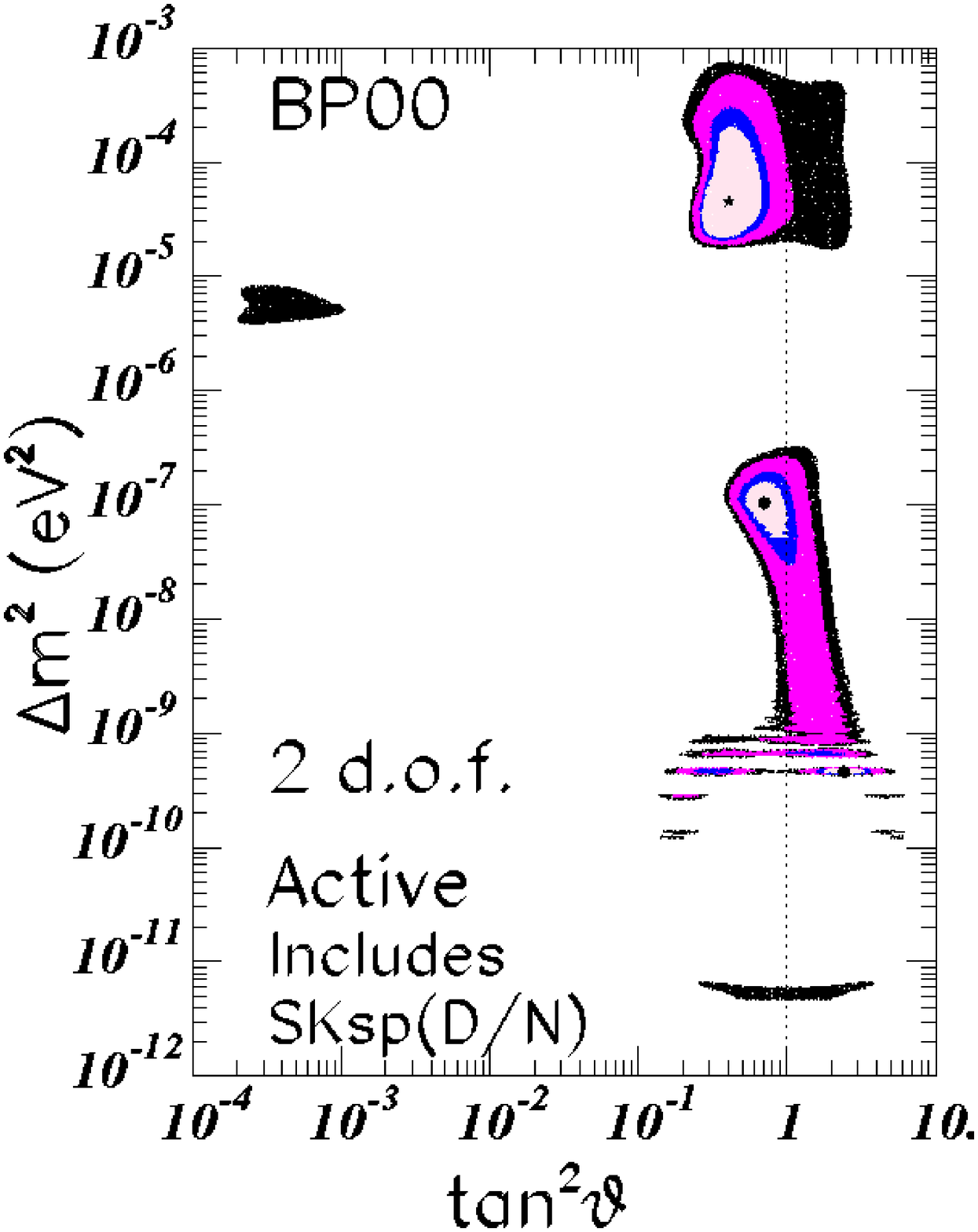,width=2.6in}\psfig{figure=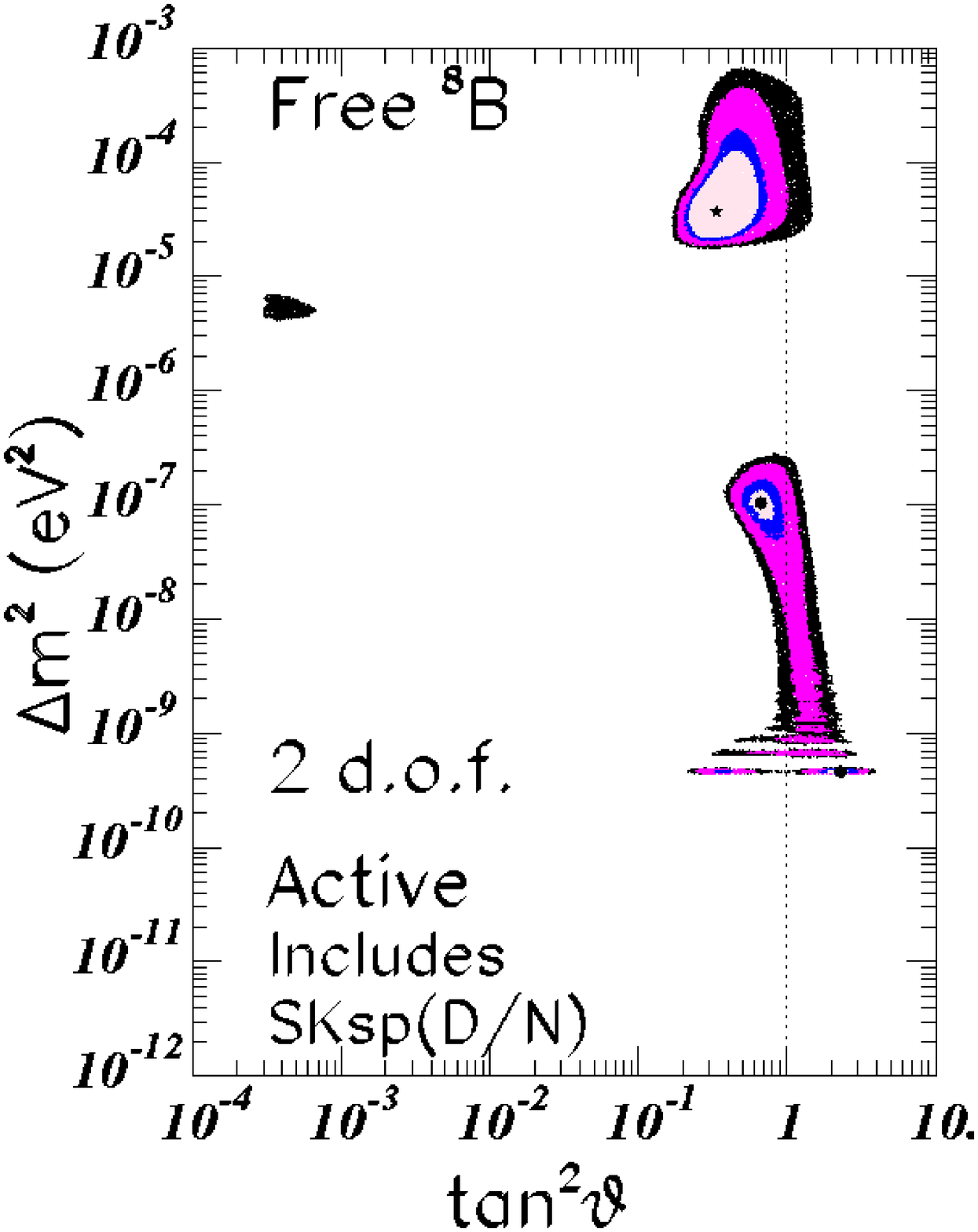,width=2.6in}}
\caption{{\bf Global solutions assuming {\it a priori} that there are
only two active neutrinos.} The input data and the analysis procedures
are the same as used in producing figure~\ref{fig:global}, except that
we have assumed for figure~\ref{fig:twodof} that there are no sterile
neutrinos and therefore the analysis is carried out for two d.o.f. .
The left and right panels in figure~\ref{fig:twodof} should be
compared, respectively, with figure~\ref{fig:global} and with
figure~\ref{fig:globalfree} of this paper and with figure~6
of ref. ~\cite{foglipostsno}.
\label{fig:twodof}} }

\subsection{Delicate differences in the error analysis}
\label{subsec:delicate}

A variation on our standard analysis scheme (the results of which are
illustrated by figure~\ref{fig:global}) can teach us a lot about the
existence or non-existence of the SMA solution.  Depending on the way
the normalization of the Super-Kamiokande total rate and the
theoretical solar model errors are included in the analysis, the
relative quality of the fits can be affected slightly for both the
active and the sterile SMA solutions as compared to the fit obtained
with the active LMA solution.

If, as done in ref.~\cite{foglipostsno}, we include in the analysis
the Super-Kamiokande total event rate with its corresponding
theoretical error, but allow a free normalization for the
Super-Kamiokande day and night energy spectra, (for which no
theoretical error is included), we obtain a worse g.o.f. for the SMA
solutions while the g.o.f. of the LMA solution is essentially
unchanged ($\chi^2_{\rm LMA}=35.5$, $\chi^2_{\rm SMA}=49.8$,
$\chi^2_{\rm sterile\,\, SMA}=53.4$).  Thus the active and sterile SMA
solutions do not appear at the $3\sigma$ level in such an analysis,
which is in agreement with the results of ref.~\cite{foglipostsno}.

We trace the origin of the difference in the presence or absence of
the SMA solution to the effect of the ``energy independent'' $^8$B
neutrino flux normalization error taken from the standard solar model
estimate~\cite{bp2000}.  As stated above, in ref.~\cite{foglipostsno},
the solar model error for $^8$B is included for the total
Super-Kamiokande rate but not for the free normalization factor in the
energy spectrum of the electron recoils.  We have included in the
present paper the effect of the BP00 errors on the predicted rate in
each energy bin of the recoil electron spectrum. When introduced in
each of the spectrum bins, this theoretical error in the total flux
produces an effectively energy-dependent error if the survival
probability depends upon energy and our procedure is therefore not
equivalent to considering the theoretical error to apply only to the
total rate.  In other words, if the expected number of events is
different in each energy bin, the theoretical error of this expected
number of events is different in each energy bin, even if the relative
error is energy independent. As a consequence, shifting up or down the
theoretical prediction within the 1$\sigma$ theoretical error, results
also in a change of the predicted spectrum shape. In particular, for a
given point in the SMA region, moving 1$\sigma$ down within the
theoretical errors (needed to fit better the observed number of events
in Super-Kamiokande and SNO) results in a flatter predicted electron
energy spectrum (less distortion due to oscillations) and fits better
the observed Super-Kamiokande recoil electron energy spectrum.

Including the $^8$B neutrino flux normalization error in the predicted
rate in each energy bin of the recoil electron energy spectrum leads
to a better fit for the Just So$^2$ solution than when the analysis is
performed with the theoretical error included only in the total
Super-Kamiokande rate (as in ref. ~\cite{foglipostsno}). For the Just
So$^2$ solution, the expected rate is larger than the measured value
at Super-Kamiokande, so the 20\% theoretical $^8$B flux normalization
error is larger for the Just So$^2$ solution than the corresponding
error for the LMA solution.  The larger error for Just So$^2$, when
added to the 38 spectrum data points, even if fully correlated, leads
to a relatively lower $\chi^2$ and also to a smaller impact of the SNO
CC measurement on the quality of the Just So$^2$ solution for the
global fit. Consistent with this explanation the Just So$^2$ solution does
not appear when the analysis is performed with the total rates only
(see, e. g., Figure~\ref{fig:ratesonly}).

We have further verified that when the analysis is performed with
unconstrained $^8$B fluxes and therefore no theoretical error for the
$^8$B flux is included, both approaches, either including the
Super-Kamiokande total event rate and a free spectrum normalization,
or not including the Super-Kamiokande total rate and using the
spectrum normalization information, lead to very similar results.
Thus the results shown in figure~\ref{fig:globalfree} hold for both of
the above-described analysis variations.

Thus the presence or absence of the SMA solution at $3\sigma$ depends
upon whether and how the standard solar model  errors on the $^8$B
neutrino flux are introduced into the analysis.

The question of whether or not the SMA solution is ruled out at
$3\sigma$ is an example where we think it is useful to apply the
motto: ``If it is not robust, it is not believable.''

\subsection{Mixtures of sterile and active neutrinos}
\label{subsec:mixtures}

We summarize in this subsection our method and results involving
mixtures of sterile and active neutrinos.  We pay particular attention
to the case in which the $^8$B neutrino flux is unconstrained. For
this case, we compare our results with those of Barger, Marfatia, and
Whisnant, ref.~\cite{barger2001}.

In order to ``level the playing field'', we treat active and sterile
neutrinos as different aspects of the same two-neutrino oscillation
scheme.  Assuming that either all solar neutrinos oscillate into all
active or all sterile neutrinos corresponds to the limiting extremes
of a continuum in which the oscillation occurs into a linear
combination of active and sterile neutrinos, cf. discussion of four
neutrino oscillations in ref.~\cite{four} .  We use the available
experimental data to determine the favored composition of the final
state. In the different figures in this paper, we show the allowed
solution regions in the limiting extremes in which the 
neutrino state into which  $\nu_e$'s oscillate is purely active
neutrinos or purely sterile neutrinos. We minimize $\chi^2$ in
3-dimensional parameter space and define the allowed regions with
respect to the global minimum in terms of shifts of $\Delta\chi^2$
for 3 d.o.f. (cf. discussion in section~\ref{subsec:defnchi2}).

For the analysis performed using the standard solar model fluxes and
errors, we find that the global best-fit point lies within the LMA
allowed region for purely active neutrinos.  This result holds for
both the global analysis using all available solar neutrino data as
well as for the analysis that includes only the information on the
total rates.

For the analysis performed for unconstrained $^8$B neutrinos, the
situation is somewhat complex. Barger and his colleagues, see
ref.~\cite{barger2001}, have pointed out that even after the SNO CC
result one can obtain a good fit for primarily oscillations into
sterile neutrinos if the $^8$B solar neutrino flux is allowed to be
considerably larger than the BP00 prediction.

We confirm the conclusion of Barger et al. on the viability of
primarily sterile neutrino oscillation solutions provided that the
analysis is performed using only the information on the total event
rates. In this case, the best fit to the rate data lies in the LMA
region but with a non-zero sterile neutrino component that increases
in amplitude as one increases the assumed $^8$B neutrino normalization
factor, $f_{\rm B}$ (for the definition of $f_{\rm B}$ see
section~\ref{subsec:defnchi2}).

However, once the Super-Kamiokande day-night spectral energy data are
included in the analysis, we find that the best fit corresponds to
purely active oscillations (zero sterile neutrino component). 
The small day-night effect measured by Super-Kamiokande is the reason
for this difference between the potentially large sterile component
that can be present in the rates-only analysis and the depressed
sterile component that is present in the global analysis.  The
solutions with a large sterile component imply a large day-night
effect in Super-Kamiokande, which was not measured.
Thus we conclude that the results of
ref.~\cite{barger2001} on the continued viability of a large sterile
solar neutrino flux, although correct for the analysis of only the total
rates, do not apply once the day-night spectral data are included.

\section{Discussion}
\label{sec:discussion}

The theme of this paper is that robust conclusions regarding solar
neutrino oscillation parameters are independent of plausible
variations in the analysis strategy. Different groups have explored
slightly different strategies and we have considered and compared in
this paper a variety of possible approaches. We show, for example,
that some conclusions depend upon the specific way that the
theoretical uncertainties are included in the analysis. We advocate
accepting as established only those inferences that do not depend
sensitively upon the details of how the analysis is performed.

Our principal conclusions are simple to state.

First, for every one of the analysis strategies used in this paper, the
favored solutions all involve large mixing angles: LMA, LOW, or
VAC. The MSW solutions are favored over the vacuum solutions if all of
the data, not just the total rates, are considered.

Second, all eight of the previously recognized two component
oscillation solutions are still allowed at $3 \sigma$ if one considers
in the analysis the solar model constraints on the $^8$B neutrino flux
and all the available neutrino data, including the SNO CC
measurement and the Super-Kamiokande spectral energy distributions in
the day and at night.  The goodness of fit for all of these solutions
ranges from good to satisfactory. The results are shown in
figure~\ref{fig:global} and table~\ref{tab:globalbestbp00}, which
describe the currently preferred global solution. 

Third, our results agree in broad outline with the recent contemporary
analyses of the solar neutrino data presented in
refs.~\cite{barger2001,foglipostsno}.  The small differences between
the separate analyses depend in physically understandable ways upon
different strategies and assumptions that have been adopted, as
discussed in section~\ref{sec:comparisons}.

Fourth, even the extreme assumption of ignoring all the
Super-Kamiokande data on the spectral recoil energy distribution and
the day-night variations does not change very much the allowed regions
of the preferred solutions. Only the marginally allowed Just So$^2$
active and sterile solutions and the SMA sterile solution are
not allowed at $3\sigma$ when the spectral and day-night data are
ignored. This result can be seen by comparing
figure~\ref{fig:ratesonly} with figure~\ref{fig:global}.

Fifth, the oscillation scenarios with a very large sterile neutrino
component that were found in ref.~\cite{barger2001} when analyzing
only the total event rates are disfavored at $3\sigma$ when the
day-night spectral energy distribution measured by Super-Kamiokande is
included in the analysis(see section~\ref{subsec:mixtures}).

Sixth, the global solutions that are obtained with and without solar
model constraints on the $^8$B neutrino flux are very similar for the
favored large mixing angle solutions. This result can be seen by
comparing visually figure~\ref{fig:global} and
figure~\ref{fig:globalfree} (both constructed using all the available
solar neutrino data) and figure~\ref{fig:ratesonly} and
figure~\ref{fig:ratb8free} (both constructed using only the data on
the total rates). The only apparent differences occur for marginally
allowed regions like the sterile solutions and the active Just So$^2$
solution. The reason for the insensitivity to how the $^8$B neutrino
flux is treated is that the combined SNO and Super-Kamiokande
measurement of the total flux~\cite{sno2001} is in excellent agreement
with the standard solar model prediction~\cite{bp2000}.

Seventh, the allowed solution space is not qualitatively affected by a
suggested~\cite{beacom} increase by $4$\% of the CC cross section on
deuterium, although the allowed regions of the SMA and the Just So$^2$
solutions become somewhat smaller when the cross section is
enhanced. This result can be inferred from a comparison of
figure~\ref{fig:global} and figure~\ref{fig:globaln}.

Eighth, we have investigated the effect of reducing the total error by
a factor of three on the experimental measurement of the $\nu_e$ 
flux while the best-estimate flux remains constant at the value quoted
in ref.~\cite{sno2001}.  The currently allowed LMA and SMA solutions
are not much affected by this hypothetical and optimistic error
reduction, but the SMA is eliminated at $3\sigma$ from the global
solution with all the data included, as are also the active and
sterile Just So$^2$ solutions. These results are apparent when
comparing figure~\ref{fig:ccdividedby3} with figure~\ref{fig:global}.

In summary, the CC measurement by SNO has not changed qualitatively
the globally allowed solution space for solar neutrinos, although the
CC measurement has provided dramatic and convincing evidence for
neutrino oscillations and has strengthened the case for active
oscillations with large mixing angles. These results are robust and
does not depend sensitively on the details of the analysis
assumptions.  Future SNO measurements~\cite{sno}, including the
day-night effect, the spectral energy distribution, and the neutral
current to charge current ratio, will significantly reduce the allowed
regions of parameter space~\cite{snoten}.  The KamLAND~\cite{kamland}
and BOREXINO~\cite{borexino} experiments will provide stringent
diagnostics of different oscillation scenarios.

We are grateful to a number of colleagues for insightful remarks and
 stimulating questions regarding an earlier draft of this paper. We
 especially appreciate comments by J. Beacom, M. Chen, R. Gandhi,
 P. Krastev, E. Lisi, A. McDonald, S. Petcov, H. Robertson, and
 G. Steigman. JNB acknowledges support from NSF grant No. PHY0070928.
 MCG-G thanks the School of Natural Sciences in the Institute for
 Advanced Study (Princeton), where part of this work was carried out,
 for warm hospitality.  CPG thanks the CERN theory division for their
 hospitality.  MCG-G is supported by the European Union Marie-Curie
 fellowship HPMF-CT-2000-00516.  This work was also supported by the
 Spanish DGICYT under grants PB98-0693 and PB97-1261, by the
 Generalitat Valenciana under grant GV99-3-1-01, and by the TMR
 network grant ERBFMRXCT960090 of the European Union and ESF network
 86.

\end{document}